\documentclass[showpacs,twocolumn,prl,aps, superscriptaddress]{revtex4}
\usepackage{amssymb}
\usepackage{amsmath}
\usepackage{graphicx}

\begin{document}

\title{Competition between weak localization and antilocalization in topological
surface states}
\author{Hai-Zhou Lu}
\affiliation{Department of Physics and Centre of Theoretical and Computational Physics,
The University of Hong Kong, Pokfulam Road, Hong Kong, China}
\author{Junren Shi}
\affiliation{International Center for Quantum Materials, Peking University, Beijing
100871, China}
\author{Shun-Qing Shen}
\affiliation{Department of Physics and Centre of Theoretical and Computational Physics,
The University of Hong Kong, Pokfulam Road, Hong Kong, China}
\date{\today }

\begin{abstract}
A magnetoconductivity formula is presented for the surface states of a
magnetically doped topological insulator. It reveals a competing effect of
weak localization and weak antilocalization in quantum transport when an energy gap is opened at the Dirac point by magnetic doping.
It is found that, while random magnetic scattering always drives the system from the
symplectic to the unitary class, the gap could induce a crossover from weak antilocalization to weak
localization, tunable by the Fermi energy or the gap. This crossover presents a unique feature characterizing the surface states of a topological insulator with the gap opened at the Dirac point in the quantum diffusion regime.
\end{abstract}

\maketitle

Topological surface states, composed of an odd number of
massless Dirac cones, are peculiar to three-dimensional (3D) topological
insulators \cite{Hasan10rmp,Qi10rmp,Moore10nat}. Electrons in these states
have a helical spin structure in momentum space, and acquire a $\pi $ Berry's
phase after completing a closed trajectory adiabatically around the Fermi
surface.
The $\pi $ Berry phase could lead to the absence of backscattering\cite{Ando98jpsj}, weak antilocalization\cite{Suzuura02prl},
and the absence of Anderson localization\cite{Bardarson07prl,Nomura07prl}.
In the quantum diffusion regime (mean free path $\ll$ system size $\sim$ phase coherent length),
an electron maintains its phase coherence after being scattered by static centers for many times.
As a result, the destructive interference due to
the $\pi$ Berry phase can give a quantum enhancement to the classical
electronic conductivity, leading to weak antilocalization (WAL)\cite{Bergmann84PhysRep,Lee85rmp}.
Applying a magnetic field tends to break the destructive interference, giving rise to negative magnetoconductivity (MC), a key signature of
WAL. WAL is expected in systems with symplectic symmetry.
Much effort has been devoted to observing WAL in graphene~\cite{Wu07prl,Gorabchev07prl,Suzuura02prl,McCann06prl,Tikhonenko09prl}.
However, graphene has two valleys of gapless Dirac cones with opposite chiralities, and the
intervalley scattering will inevitably suppress WAL~\cite%
{Suzuura02prl,Wu07prl,Gorabchev07prl,McCann06prl,Tikhonenko09prl}. In
contrast, the surface states of recently discovered topological insulators Bi$_{2}$Te$_{3}$
and Bi$_{2}$Se$_{3}$ have only one helical Dirac cone \cite{Xia09np,Zhang09np,Chen09science},
and WAL is intrinsic to them.
Many observations of WAL in Bi$_{2}$Te$_{3}$ and Bi$_{2}$Se$_{3}$ have been
reported recently \cite{Peng10NatMat,Chen10prl,Checkelsky10arxiv,He10arxiv,Liu10arxiv,Wang10arxiv}.
In particular, there is great interest in the effect of magnetic doping, which is considered to be an efficient way to open
an energy gap in the Dirac cone by breaking time reversal symmetry (TRS) \cite{Hor10prb,Chen10sci,Wray2011np}.
This gap is expected to give rise to many
interesting phenomena, such as Majorana fermion\cite{Majorana}, topological magnetoelectric effect \cite{Qi08prb} and
quantized anomalous Hall effect \cite{Yu10sci}. These developments call for a thorough
theoretical investigation on WAL in topological insulators, in particular,
in the presence of magnetic doping.

\begin{figure}[tbph]
\centering \includegraphics[width=0.45\textwidth]{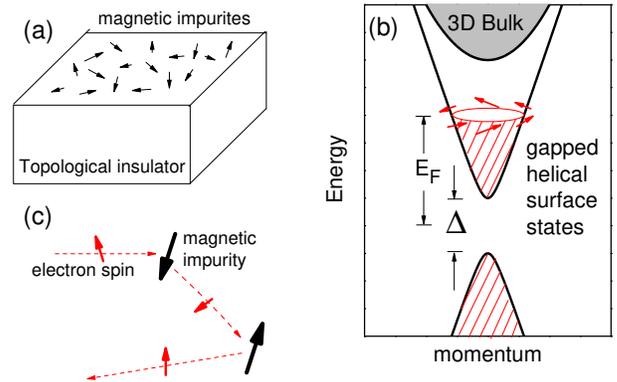}
\caption{(a) A topological insulator with magnetic doping on the
top surface. (b) Magnetic doping may open a gap ($\Delta$) at the
Dirac point of the surface states \cite{Chen10sci,Wray2011np}. $E_F$ is the Fermi energy measured from the Dirac point.
$\Delta$ tilts in-plane spin polarization of the massless Dirac fermion out of
plane, leading to the deviation from the $\pi$ Berry phase. (c) Scattering of an electron by random magnetic
impurities. Dashed lines represent the trajectory of the electron.}
\label{fig:bands}
\end{figure}

In this Letter, a MC formula is presented for the magnetically doped surface states of
a topological insulator [Fig. \ref{fig:bands}(a)].
We assume that the mean field produced by magnetic doping may open a uniform gap at the Dirac point\cite{Chen10sci,Wray2011np} [Fig. \ref{fig:bands}(b)], and the local fluctuation over the mean field can scatter
conducting electrons in a random fashion [Fig. \ref{fig:bands}(c)].
With the help of the diagrammatic technique \cite{Bergmann84PhysRep,Lee85rmp,HLN80,Altshuler80prb,Iordanskii94jetp,Shon98jpsj,Suzuura02prl,McCann06prl,Yan08prl,Imura09prb,Imura10epl},
we obtain the MC formula, which consists of two competing terms. Besides the WAL term due to the gapless Dirac fermion, an extra weak localization (WL) term arises as a result of the gap opening. We find that either the gap or the magnetic scattering can drive MC of the
system from WAL
to a parabolic dependence on the magnetic field ($\sim B^2$). Further increasing the gap/Fermi energy ratio may drive the
system to WL after reaching the $B^2$ regime. A crossover from WAL to WL is thus expected, tunable by the size of the TRS-breaking
gap or the position of Fermi energy. Beyond the theories for non-Dirac systems with strong spin-orbit coupling\cite{Iordanskii94jetp} or strong ferromagnetism\cite{Dugaev01prb},
the observation of the crossover will provide an important signature in transport experiments for the existence of a TRS-breaking gap in the topological surface
states.

We describe the magnetically doped surface states of a 3D topological insulator by the massive Dirac model $H$
in a random impurity potential $U(\mathbf{r})$. The Hamiltonian of the massive Dirac model is given by
\begin{equation}
H=\gamma (\sigma _{x}k_{y}-\sigma _{y}k_{x})+h_{z}\sigma _{z},
\end{equation}%
where $\sigma _{x,y,z}$ are the Pauli matrices. $\gamma =\hbar v_{\mathrm{F}%
} $, with $v_{\mathrm{F}}$ the Fermi velocity. The $\sigma _{z}$ term represents a
gap opened by breaking TRS, with $h_{z}=\frac{1}{2}g\mu _{B}B+\frac{\Delta }{2}
$, where the first term is the Zeeman energy of the out-of-plane magnetic
field $B$, with $g$ the g-factor and $\mu _{B}$ the Bohr magneton. $\Delta $
is the gap opened at the Dirac point \cite{Chen10sci,Wray2011np}, it originates from the mean field produced by the magnetic doping. The
Hamiltonian describes two energy bands as shown in Fig. \ref{fig:bands}(b).
In this work, we assume that the Fermi energy $E_{F}$ is tuned into the gap
of the 3D bulk bands, and intersects with the upper band of the
surface states, which is the only band included in the calculation. Its band
dispersion is given by $\epsilon _{\mathbf{k}}=\sqrt{\gamma
^{2}k^{2}+h_{z}^{2}}$ and the wave function $\psi _{\mathbf{k}}(\mathbf{r}%
)=[a,-ie^{i\varphi }b]^{T}e^{i\mathbf{k}\cdot \mathbf{r}}/\sqrt{S}$, where $%
\tan \varphi \equiv k_{y}/k_{x}$, $\mathbf{k}$ is the wave vector, $a\equiv
\cos \frac{\theta }{2}$, $b\equiv \sin \frac{\theta }{2}$, and $\cos \theta
\equiv h_{z}/\sqrt{h_{z}^{2}+\gamma ^{2}k^{2}}$, $S$ is the area.
In this work, all the physical quantities will be evaluated at the Fermi energy $E_F$ at low temperatures.
The density of states at $E_{F}$ is $N_{F}=E_{F}/(2\pi \gamma
^{2})$. The scattering by nonmagnetic and magnetic impurities is
modeled by the random potential
\begin{equation}
U(\mathbf{r})=\sum_{i,\alpha} u_{\alpha}^{i}\sigma_{\alpha}\delta (\mathbf{r}-\mathbf{R}_i),
\end{equation}
where $\alpha$ runs over $0,x,y,z$. $\sigma_{0}$ is the $2\times 2$ unit matrix.
$\mathbf{R}_{i}$ are the positions of the randomly distributed impurities.
$u_{0}^{i}$ depicts the potential at $\mathbf{R}_{i}$ for nonmagnetic
impurity, and $u_{x,y,z}^{i}$ for magnetic impurity.
Note that $u_{x,y,z}^i$ do not represent the total local exchange field produced by the impurity at $\mathbf{R}_i$, but the local fluctuation over the mean field that gives the gap.
Therefore, $\langle U(\mathbf{r})\rangle_{\mathrm{imp}
}=0$ and we can still assume the random potential is delta-correlated $\langle U(\mathbf{r})U(\mathbf{r^{\prime }})\rangle _{\mathrm{imp}}\sim \delta (\mathbf{r}-\mathbf{r}^{\prime })$, where $\langle ...\rangle _{\mathrm{imp}}$ means average over impurity configurations, and we follow the
practical assumption that different types of impurity scattering are
uncorrelated~\cite{McCann06prl}.

\begin{figure}[tbph]
\centering \includegraphics[width=0.4\textwidth]{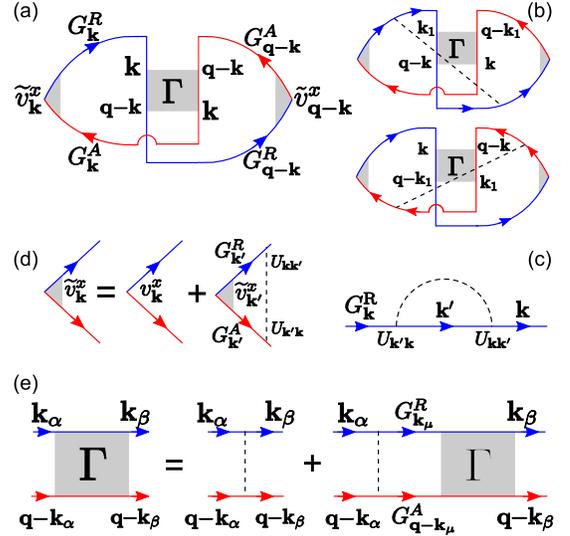}
\caption{The diagrams for the quantum interference correction to conductivity of Dirac
fermions. The arrowed solid and dashed lines represent the Green's functions and
impurity scattering, respectively. (a) The bare \cite{Altshuler80prb,HLN80} and (b) two dressed \cite{McCann06prl} Hikami
boxes give the quantum conductivity correction from the maximally crossed
diagrams. (c) The retarded Green function with the first-order Born approximation to
the impurity-averaged self-energy. (d) Ladder diagram vertex correction to
velocity\protect\cite{Shon98jpsj}. (e) The Bethe-Salpeter equation for the
vertex of maximally crossed diagrams. }
\label{fig:diagram}
\end{figure}

The quantum interference correction to conductivity of Dirac fermions can be calculated
by the diagrams in Fig. \ref{fig:diagram}, which are different from those
for the usual two-dimensional electron gas (2DEG) \cite{Bergmann84PhysRep,Lee85rmp,HLN80,Altshuler80prb} in several aspects
\cite{Shon98jpsj,Suzuura02prl,McCann06prl}. (i) Besides the conventional
maximally crossed diagram (bare Hikami box) in Fig. \ref{fig:diagram}(a),
two dressed Hikami boxes in Fig. \ref{fig:diagram}(b) are also needed; each
gives $-1/4$ as the bare Hikami box for the gapless Dirac cone. (ii) The
ladder diagram correction to the bare velocity $v_{\mathbf{k}}^{x}\equiv
(1/\hbar )\partial \epsilon _{\mathbf{k}}/\partial k_{x}$ [Fig. \ref{fig:diagram}(d)] must be taken into account, which corrects the velocity to $
\widetilde{v}_{\mathbf{k}}^{x}=2v_{\mathbf{k}}^{x}$ for the gapless Dirac
cone. We generalize these conclusions for the gapless case to the gapped
Dirac cone as follows.

The arrowed lines in Fig. \ref{fig:diagram} stand for the impurity-averaged
retarded (R) and advanced (A) Green's functions $G_{\mathbf{k}}^{\mathrm{%
\mathrm{R/A}}}(\omega )=1/(\omega -\epsilon _{\mathbf{k}}\pm i\hbar /2\tau )$%
, where under the first-order Born approximation, the impurity-induced self-energy is given
by the total scattering time $\tau $, with $\hbar /\tau \equiv 2\pi \sum_{%
\mathbf{k}^{\prime }}\langle |U_{\mathbf{k}^{\prime }\mathbf{k}}|^{2}\rangle
_{\mathrm{imp}}\delta (\omega -\epsilon _{\mathbf{k}^{\prime }})$, where $U_{%
\mathbf{k},\mathbf{k}^{\prime }}\equiv \langle \psi _{\mathbf{k}}(\mathbf{r}%
)|U(\mathbf{r})|\psi _{\mathbf{k}^{\prime }}(\mathbf{r})\rangle $ is the
scattering amplitude between two momenta. It can be separated into $1/\tau
=1/\tau _{e}+1/\tau _{m}$, where the nonmagnetic elastic scattering time $%
\tau _{e}$ is given by $\hbar /\tau _{e}=2\pi
N_{F}n_{0}u_{0}^{2}(a^{4}+b^{4})$ and the total magnetic scattering time $%
\tau _{m}$ can be separated into $1/\tau _{m}=2/\tau _{x}+1/\tau _{z}$, with
$\hbar /\tau _{z}=2\pi N_{F}n_{m}u_{z}^{2}(a^{4}+b^{4})$ and $\hbar /\tau
_{x}=2\pi N_{F}n_{m}u_{x}^{2}(2a^{2}b^{2})$. In-plane isotropy ($u_{x}=u_{y}$) is assumed.
$u_{0}$ depicts the average scattering
strength for nonmagnetic impurities, while $u_{x,y,z}$ for magnetic
impurities. $n_{0}$ and $n_{m}$ are concentrations of nonmagnetic and
magnetic impurities, respectively. $\tau _{e}$ and $\tau _{m}$ are related
to the elastic scattering length $\ell _{e}$ and magnetic scattering length $%
\ell _{m}$ by $\ell _{e}=\sqrt{D\tau _{e}}$ and $\ell _{m}=\sqrt{D\tau _{m}}$%
, respectively. $D\equiv v_F^2\tau^2/2$ is the diffusion constant. Considering the poor surface
mobility\cite{Checkelsky10arxiv,Chen10prl,He10arxiv}, we assume that $\ell _{e}$
is much shorter than the phase coherence length $\ell _{\phi }$, as required
by the quantum diffusion transport.

In the conductivity diagrams Figs. \ref{fig:diagram}(a) and (b), the
vertex function $\Gamma $ from the maximally crossed diagrams usually
is proportional to $1/q^{2}$, where $\mathbf{q}$ is the
summation of momenta before and after scattering.
Since $\Gamma$ diverges as $q\rightarrow 0$, it contributes
mainly to backscattering. This allows us to sum $\mathbf{k}$ and $\mathbf{k}%
_{1}$ for small $q$ first, and write the zero-temperature conductivity
correction from the bare and two dressed Hikami boxes as
\begin{equation}\label{sigma_F}
\sigma ^{F}=-\frac{e^{2}N_{F}v_{F}^{2}\tau ^{3}\sin ^{2}\theta}{\hbar ^{2}}\eta
_{v}^{2} (1+2\eta _{H})\sum_{\mathbf{q}}\Gamma (\mathbf{q}),
\end{equation}
where $\eta _{v}$ comes from the correction to velocity from the ladder diagrams
in Fig. \ref{fig:diagram}(d), with
\begin{equation}
\eta _{v}=\left[ 1-\frac{1}{2}(\frac{\tau }{\tau _{e}}-\frac{\tau }{\tau _{z}%
})\frac{2a^{2}b^{2}}{a^{4}+b^{4}}\right] ^{-1},
\end{equation}
and each of the dressed Hikami boxes gives an extra $\eta_H$ contribution as the
bare Hikami box, with
\begin{equation}
\eta _{H}=-\frac{1}{2}\left( 1-\eta _{v}^{-1}-\frac{\tau }{\tau_{x}}
\right).
\end{equation}
$\eta_{v}$ and $\eta_{H}$ reduce to $2$ and $-1/4$, respectively, in the
absence of magnetic doping \cite{Shon98jpsj,Suzuura02prl,McCann06prl}.

In Fig. \ref{fig:diagram}, the total momentum conserves on the incoming and
outgoing sides of the vertex $\Gamma $, allowing the Bethe-Salpeter
equation of the vertex to be written as \cite{Suzuura02prl}
$\Gamma _{\mathbf{k}_{\alpha }\mathbf{k}_{\beta }}=\Gamma _{\mathbf{k}%
_{\alpha }\mathbf{k}_{\beta }}^{0}+\sum_{\mathbf{k}_{\mu }}\Gamma _{\mathbf{k%
}_{\alpha }\mathbf{k}_{\mu }}^{0}G_{\mathbf{k}_{\mu }}^{R}G_{\mathbf{q}-%
\mathbf{k}_{\mu }}^{A}\Gamma _{\mathbf{k}_{\mu }\mathbf{k}_{\beta }},
$ where $\mathbf{k}_{\alpha }+\mathbf{k}_{\beta }=\mathbf{q}$, $\mathbf{k}%
_{\alpha ,\beta }$ are the incoming and outgoing momenta, respectively. For
small $q$, the bare vertex $\Gamma _{\mathbf{k}_{\alpha }\mathbf{k}_{\beta
}}^{0}\equiv \langle U_{\mathbf{k}_{\beta },\mathbf{k}_{\alpha }}U_{\mathbf{q%
}-\mathbf{k}_{\beta },\mathbf{q}-\mathbf{k}_{\alpha }}\rangle _{\mathrm{imp}%
} $ is found as $\Gamma _{\mathbf{k}_{\alpha }\mathbf{k}_{\beta }}^{0}\approx \frac{\hbar}{2\pi N_F}[ A+Be^{i(\varphi
_{\alpha }-\varphi _{\beta })}+Ce^{i2(\varphi _{\alpha }-\varphi _{\beta })}]
$, with $A=(\tau _{e}^{-1}+\tau _{z}^{-1})\frac{a^{4}}{%
a^{4}+b^{4}}$, $B=[(\tau _{e}^{-1}-\tau _{z}^{-1})%
\frac{2a^{2}b^{2}}{(a^{4}+b^{4})}-2\tau _{x}^{-1}]$, $C=(\tau _{e}^{-1}+\tau _{z}^{-1})\frac{b^{4}}{a^{4}+b^{4}}$. Different
from the usual 2DEG, both the bare vertex $\Gamma _{\mathbf{k}_{\alpha }%
\mathbf{k}_{\beta }}^{0}$ and the advanced Green function $G_{\mathbf{q}-%
\mathbf{k}_{\mu }}^{A}$ are explicit functions of the momentum angle. We propose
an ansatz to the full vertex function
\begin{equation}
\Gamma _{\mathbf{k}_{\alpha }\mathbf{k}_{\beta }}=\frac{\hbar }{2\pi
N_{F}\tau }\sum_{n,m\in 0,1,2}\gamma _{nm}e^{i(n\varphi _{\alpha }-m\varphi
_{\beta })},
\end{equation}
where $\gamma _{nm}$ are the expansion coefficients independent of $\varphi
_{\alpha ,\beta }$. By putting the ansatz into the Bethe-Salpeter equation and
expanding $G_{\mathbf{q}-\mathbf{k}_{\mu }}^{A}$ up to $q^{2}$, we obtain the solution
to the expansion coefficients
\begin{equation}
\boldsymbol{\gamma}=2\left[
\begin{array}{ccc}
g_{0}+Q^{2} & iQ_{+} & \frac{1}{2}Q_{+}^{2} \\
iQ_{-} & g_{1}+Q^{2} & iQ_{+} \\
\frac{1}{2}Q_{-}^{2} & iQ_{-} & g_{2}+Q^{2}%
\end{array}%
\right] ^{-1},
\end{equation}
where $Q_{\pm }$=$Q_{x}$$\pm $$iQ_{y}$, $Q^{2}$=$Q_{x}^{2}$+$Q_{y}^{2}$,
$\mathbf{Q}=v_{F}\tau \sin \theta (q_{x},q_{y})$, and the ``Cooperon gaps" $
g_{0} \equiv 2[\frac{a^{4}+b^{4}}{a^{4}}\frac{1/\tau }{(1/\tau _{e}+1/\tau
_{z})}-1]$, $g_{1} \equiv 2[\frac{1/\tau }{(1/\tau _{e}-1/\tau _{z})\frac{2a^{2}b^{2}}{%
a^{4}+b^{4}}-2/\tau _{x}}-1]$, $g_{2} \equiv 2[\frac{a^{4}+b^{4}}{b^{4}}\frac{1/\tau }{(1/\tau _{e}+1/\tau
_{z})}-1]$.
We note that it is crucial to include all the off-diagonal terms of $\boldsymbol{\gamma}$ in the calculation.
Without them, the vertex will be 2 times larger
\cite{Imura09prb,Imura10epl} when going back to the gapless limit\cite{Suzuura02prl},
and the derived MC formula can not recover to that for 1/4 of graphene \cite{McCann06prl}.
$\Gamma (\mathbf{q})$ in Eq. (\ref{sigma_F}) can be obtained by letting $\mathbf{k}%
_{\alpha }=\mathbf{k}$ and $\mathbf{k}_{\beta }=\mathbf{q}-\mathbf{k}$ in $%
\Gamma _{\mathbf{k}_{\alpha }\mathbf{k}_{\beta }}$, and for $q \rightarrow 0$, $%
\varphi _{\mathbf{k}}-\varphi _{\mathbf{q}-\mathbf{k}}\approx \pi $.
Finally, we collect the most divergent terms of the vertex
\begin{equation}
\Gamma (\mathbf{q})\approx   \frac{\hbar /(\pi N_{F}\tau )}{g_{0}+(1+%
\frac{1}{g_{1}})Q^{2}}-\frac{\hbar /(\pi N_{F}\tau )}{g_{1}+(1+\frac{1}{g_{0}%
}+\frac{1}{g_{2}})Q^{2}}  .
\end{equation}

\begin{figure}[tbph]
\centering \includegraphics[width=0.45\textwidth]{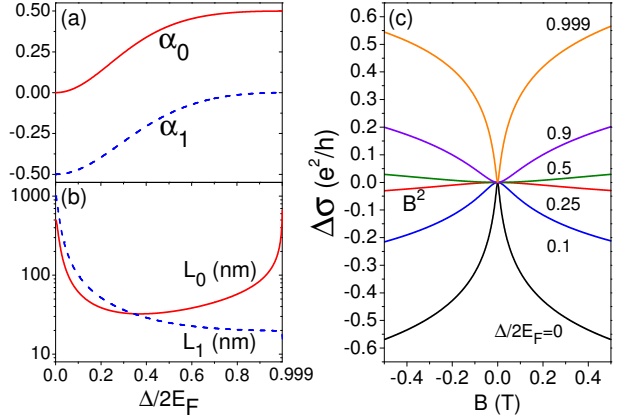}
\caption{(a) WL ($\protect\alpha_0$) and WAL ($\protect\alpha_1$) weight
factors as functions of $\Delta/2E_F$, where $\Delta$ is the gap, $E_F$ is
the Fermi energy. (b) WL ($\ell_0$) and WAL ($\ell_1$) lengths as functions
of $\Delta/2E_F$. (c) Magnetoconductivity $\Delta\protect\sigma(B)$ for
different $\Delta/2E_F$ in the limit of weak magnetic scattering. $\ell_{\protect\phi}=300$ nm. $\ell_m=1000$ nm. $%
u_x=u_z$ is assumed.}
\label{fig:mc_gap}
\end{figure}

Zero-field conductivity correction $\sigma ^{F}(0)$ can be calculated by
performing the integral over $q$ in Eq. (\ref{sigma_F}) between $1/\ell
_{e} $ and $1/\ell _{\phi }$, respectively~\cite{Suzuura02prl}. In the
presence of the perpendicular magnetic field $B$, $q^{2}$ will be quantized
into $q_{n}^{2}=(n+1/2)(4eB/\hbar )\equiv (n+1/2)/\ell _{B}^{2}$, where $n$ labels the
Landau levels. Summation over $n$ gives the conductivity correction $\sigma
^{F}(B)$ at finite field \cite{Bergmann84PhysRep}. The magnetoconductivity $\Delta \sigma (B)$$\equiv
\sigma ^{F}(B)-\sigma ^{F}(0)$ is found for $\ell _{B}^{2}/\ell
_{e}^{2}\gg 1$ as
\begin{equation}
\Delta \sigma (B)=\sum_{i=0,1}\frac{\alpha _{i}e^{2}}{\pi h}\left[ \Psi (%
\frac{\ell _{B}^{2}}{\ell _{\phi }^{2}}+\frac{\ell _{B}^{2}}{\ell _{i}^{2}}+%
\frac{1}{2})-\ln (\frac{\ell _{B}^{2}}{\ell _{\phi }^{2}}+\frac{\ell _{B}^{2}%
}{\ell _{i}^{2}})\right] ,
\end{equation}
with $\Psi $ the digamma function,
\begin{eqnarray}
&&\alpha _{1}=-\frac{\eta _{v}^{2}(1+2\eta _{H})}{2(1+\frac{1}{g_{0}}+\frac{1%
}{g_{2}})},\ \ell _{1}^{-2}=\frac{g_{1}}{2 \ell^2 \sin ^{2}\theta
(1+\frac{1}{g_{0}}+\frac{1}{g_{2}})},  \notag \\
&&\alpha _{0}=\frac{\eta _{v}^{2}(1+2\eta _{H})}{2(\frac{1}{g_{1}}+1)},\
\ell _{0}^{-2}=\frac{g_{0}}{2\ell^2 \sin ^{2}\theta (\frac{1}{g_{1}%
}+1)},
\end{eqnarray}
and $1/\ell^{2}\equiv 1/\ell^2_e+1/\ell^2_m $.
In the absence of magnetic impurities, $\alpha _{0}=0$, $\alpha _{1}=-1/2$, one
predicts WAL with a prefactor $-1/2$, consistent with the experimental
observations \cite{Checkelsky10arxiv,Chen10prl,He10arxiv}. For a finite gap,
because $\alpha _{0}$ and $\alpha _{1}$ have opposite signs, the MC formula
has two competing contributions, $\alpha _{1}$ leads to WAL, $\alpha _{0}$
to WL. $\ell _{0}$ and $\ell _{1}$ give corrections to $\ell _{\phi }$, in
particular, when they are much shorter than $\ell _{\phi }$. This formula is
the key result of this work.

\begin{figure}[tbph]
\centering \includegraphics[width=0.45\textwidth]{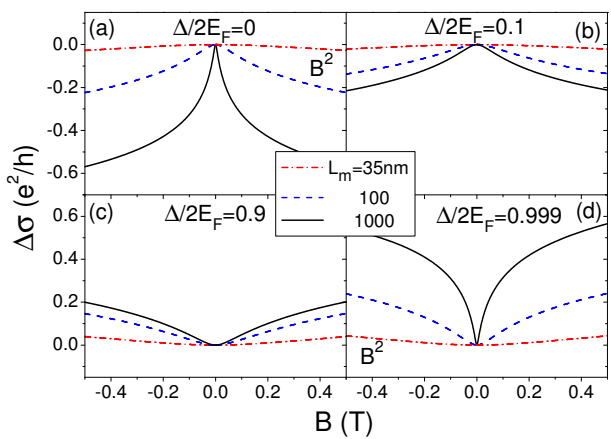}
\caption{Magnetoconductivity $\Delta\protect\sigma(B)$ for different
magnetic scattering lengths $\ell_m$ and $\Delta/2E_F$. $\ell_{\protect\phi%
}=300$ nm. $u_x=u_z$ is assumed. Shorter $\ell_m$ means stronger magnetic scattering.}
\label{fig:mc_Lm}
\end{figure}

We first examine the limit of weak magnetic scattering, i.e., $\ell _{m}\gg \ell
_{\phi }$~\cite{Peng10NatMat,Checkelsky10arxiv,Chen10prl,He10arxiv}. We plot
MC for different $\Delta /2E_{F}$ in Fig. \ref{fig:mc_gap}(c). For $\Delta
/2E_{F}=0$, MC shows a positive cusp, which is the signature of WAL. As $%
\Delta /2E_{F}$ increases, MC gradually develops a $B^{2}$
dependence, and the system evolves into the unitary regime. Further
increasing $\Delta /2E_{F}$ will change the sign of MC from negative to
positive, i.e., a WL-like MC. Different from the usual 2DEG, the WL-like MC
here has a prefactor $\sim 1/2$, instead of $1$~\cite{HLN80}. This can be
seen in Fig. \ref{fig:mc_gap}(a), where we show the weight factors of
the competing WL and WAL terms in the MC formula. In the limit of small $%
\Delta /2E_{F}$, $\alpha _{1}$ overweighs $\alpha _{0}$, so MC is mainly
contributed by WAL, with the maximal prefactor $-1/2$. In the limit of large
$\Delta /2E_{F}$, $\alpha _{1}$ vanishes and $\alpha _{0}$ goes to $1/2$,
then we have WL with the maximal prefactor 1/2. As shown in Fig. \ref{fig:mc_gap}(b),
either $\ell _{1}$ for small $\Delta /2E_{F}$ or $\ell _{0}$ for large $%
\Delta /2E_{F}$ is much larger than $\ell _{\phi }$, this keeps the system
well inside the quantum diffusion regime, and protects WAL or WL. For
intermediate $\Delta /2E_{F}$, where both the WL and WAL terms contribute,
the weak $B^{2}$ MC indicates that the system is driven from the quantum to
classical diffusion regime due to the effective reduction of $\ell _{\phi }$
by the much shorter $\ell _{0}$ and $\ell _{1}$. The crossover from WAL to
WL by changing $\Delta /2E_{F}$ can be understood with the Berry phase\cite{Ando98jpsj},
which is readily evaluated for the surface band $\psi _{\mathbf{k}}(\mathbf{r%
})$ as
\begin{equation}
-i\int_{0}^{2\pi }d\varphi \left\langle \psi _{\mathbf{k}}(\mathbf{r}%
)\left\vert \frac{\partial }{\partial \varphi }\psi _{\mathbf{k}}(\mathbf{r}%
)\right. \right\rangle =\pi (1+\frac{\Delta }{2E_{F}}).
\end{equation}
It gives $\pi $ for WAL when $\Delta /2E_{F}= 0$, and $2\pi $ for
WL when $\Delta =2E_{F}$.
A similar argument was also given for the gap opened by the finite size effect \cite{Ghaemi10prl}.
In the limit of strong magnetic scattering $\ell _{m}\ll \ell _{\phi }$,
both WAL and WL are suppressed, as shown in Fig. \ref{fig:mc_Lm}. On the
other hand, because the Fermi energy in the ratio $\Delta /E_{F}$ can be
controlled independently by gate voltage~\cite{Checkelsky10arxiv,Chen10prl},
it is possible to observe the transition from negative to positive MC by
tuning the gate voltage even in this limit.

We thank H. T. He, J. N. Wang, F. C. Zhang, W. Q. Chen, X. Dai, and B. Zhou
for helpful discussions. Hai-Zhou also thank K. Imura and H. Suzuura for
stimulating discussions. This work is supported by the Research Grant
Council of Hong Kong under Grant No. HKU 7051/10P and HKUST3/CRF/09. Junren is supported
by NSFC No. 10734110 and 973 program of China No. 2009CB929101.

\end{document}